\documentclass[12pt]{iopart}
\usepackage{harvard}
\usepackage{psfig}
\begin{document}

\article[Status Update of the PPTA]{Proceedings of the 8th Amaldi
  Conference}{Status Update of the Parkes Pulsar Timing Array}

\author{J P W Verbiest$^1$, M Bailes$^2$, N D R Bhat$^2$, S
  Burke-Spolaor$^{2,3}$, D J Champion$^4$, W Coles$^5$, G B Hobbs$^3$,
  A W Hotan$^6$, F Jenet$^7$, J Khoo$^3$, K J Lee$^4$, A Lommen$^8$, R
  N Manchester$^3$, S Oslowski$^{2,3}$, J Reynolds$^3$, J
  Sarkissian$^3$, W van Straten$^2$, D R B Yardley$^{3,9}$ and X P
  You$^{10}$}

\address{$^1$ Department of Physics, West Virginia University,
  P.O. Box 6315, WV 26506, USA}\ead{Joris.Verbiest@mail.wvu.edu}
\address{$^2$ Swinburne University of Technology, Centre for
  Astrophysics and Supercomputing, Mail \#H39, P.O. Box 218, VIC 3122,
  Australia}
\address{$^3$ Australia Telescope National Facility -- CSIRO, P.O. Box
  76, Epping, NSW 1710, Australia}
\address{$^4$ Max-Planck-Institut f\"ur Radioastronomie, Auf dem
  H\"ugel 69, 53121 Bonn, Germany}
\address{$^5$ Electrical and Computer Engineering, University of
  California at San Diego, La Jolla, CA 92093, USA}
\address{$^6$ Curtin Institute for Radio Astronomy, Curtin University
  of Technology, Bentley, WA 6102, Australia}
\address{$^7$ CGWA, University of Texas at Brownsville, TX 78520, USA}
\address{$^8$ Franklin and Marshall College, 415 Harrisburg Pike,
  Lancaster, PA 17604, USA}
\address{$^{9}$ Sydney Institute for Astronomy, School of Physics
  A29, The University of Sydney, NSW 2006, Australia}
\address{$^{10}$ School of Physical Science and Technology, Southwest
  University, 2 Tiansheng Road, Chongqing 400715, China}

\begin{abstract}
  The Parkes Pulsar Timing Array project aims to make a direct
  detection of a gravitational-wave background through timing of
  millisecond pulsars. In this article, the main requirements for that
  endeavour are described and recent and ongoing progress is
  outlined. We demonstrate that the timing properties of millisecond
  pulsars are adequate and that technological progress is timely to
  expect a successful detection of gravitational waves within a
  decade, or alternatively to rule out all current predictions for
  gravitational wave backgrounds formed by supermassive black-hole
  mergers.
\end{abstract}

\submitto{\CQG}
\maketitle

\section{Introduction}\label{sec:Intro}
Pulsars are rapidly rotating neutron stars that emit beamed radio
waves, probably from polar magnetic field lines. Due to a misalignment
of the magnetic and rotation axes, this emission is observed as pulses
of radiation whose times-of-arrival (TOAs) at a radio observatory can
be measured accurately. Pulsar timing is a technique based on the
comparison of these observed TOAs to those predicted by a timing model
that describes the spin behaviour of the pulsar, its binary orbit, the
interstellar medium and the solar-system ephemerides. The differences
between the model-predicted TOAs and the actual observed TOAs are
called the timing residuals and they contain all the information not
properly described in the timing model, such as calibration errors,
instrumental instabilities, higher-order ISM effects, instabilities
intrinsic to the pulsar, gravitational waves (GWs) and radiometer
noise.

The highly predictable nature of pulsars makes them useful tools for a
variety of investigations in physics and astrophysics. One type of
investigation was first proposed by \citeasnoun{saz78} and
\citeasnoun{det79}, who claimed that pulsar timing might be used to
detect GWs. \citeasnoun{hd83} expanded on that work and showed that a
stochastic background of GWs (GWB) would introduce a quadrupolar
correlation in pulsar timing residuals: the residuals of pulsars with
separations on the sky close to 0$^{\circ}$ or 180$^{\circ}$ would be
positively correlated, while those of pulsars separated by
90$^{\circ}$ would be anticorrelated.
% \citeasnoun{hd83} expanded on that early work and showed that a
% stochastic background of GWs (GWB) would affect pulsar timing
% residuals in such a way that the correlation of the residuals would
% depend on the angular separation between the pulsars in a
% quadrupolar way.
This insight triggered the concept of a pulsar timing array
(PTA), proposed by \citeasnoun{rom89} and \citeasnoun{fb90}. A PTA is
an array of pulsars which are timed in order to investigate correlated
signals between the timing residuals. Three causes of correlation are
known to exist: a monopolar correlation due to reference clock errors;
a dipolar correlation due to offsets in the solar-system ephemerides;
and a quadrupolar correlation due to GWs.

In this paper we describe one of the three PTAs currently in
operation: the Parkes Pulsar Timing Array (PPTA). In \S\ref{sec:PPTA}
we describe the overall project, its aims and observational
requirements as derived by \citeasnoun{jhlm05}. \S\ref{sec:progress}
details current progress, both on scientific and technological fronts
and in \S\ref{sec:GWBSens} our sensitivity to both a GWB and
individual GW sources is discussed.

\section{The Parkes Pulsar Timing Array Project}\label{sec:PPTA}
Founded in 2004, the PPTA project 
%\citeaffixed{man08}{most fully described by} 
observes 20 millisecond pulsars (MSPs - pulsars with millisecond
periods) on a bi-weekly basis with the 64-m Parkes radio telescope in
NSW, Australia, at three different observing frequencies
(centred on 685\,MHz, 1369\,MHz and 3128\,MHz). For further technical
details see \citeasnoun{man08}.

\subsection{Aims}\label{ssec:aims}
The main aims of the PPTA project are threefold: 
\begin{enumerate}
\item Make a direct detection of GWs through pulsar timing and in
  so doing, develop the tools and methodologies required for the new
  era of full-fledged GW astronomy using pulsars, which is expected to
  commence with the commissioning of the Square Kilometre Array
  \citeaffixed{kbc+04}{SKA; }.
\item Construct a pulsar time-scale, thereby removing long-term
  dependence on Earth-based clocks, enabling further improvements in
  timing precision.
\item Improve the solar-system ephemerides, both through improving
  precision of the masses of known solar-system objects and eventually
  through detection of currently unknown trans-Neptunian objects.
\end{enumerate}

Of these aims, the one with the highest expected scientific value is
GW detection. Because the most likely source of GWs to be detected by
the PPTA is a GWB formed by super-massive black-hole mergers
\cite{jhv+06}, we will describe the observational requirements for a
successful GWB detection in the following section, based on the
semi-analytic work done by \citeasnoun{jhlm05}.

\subsection{Pulsar timing array requirements}\label{ssec:requirements}
While the original PTA described by \citeasnoun{rom89} and
\citeasnoun{fb90} only contained three pulsars, more recent work by
\citeasnoun{jhlm05} has demonstrated the strong dependence of PTA
sensitivity on the number of pulsars included in the
array. Specifically, they demonstrated that the detection significance
a PTA would achieve, can be approximated as \cite[their
Eq. 12]{jhlm05}:
\begin{equation}\label{eq:S}
  S = \sqrt{{M\left(M-1\right)/2}
               \over
               {1+\left[\chi\left(1+\bar{\xi^2}\right)
                 + 2\left(\sigma_{\rm n}/\sigma_{\rm g}\right)^2
               + \left(\sigma_{\rm n}/\sigma_{\rm g}\right)^4\right]
           /N\sigma_{\rm \xi}^2}},
\end{equation}
where $M$ is the number of pulsars in the timing array, $N$ is the
number of observations of each pulsar, $\sigma_{\rm n}$ is the RMS of
the timing residuals (assumed identical for all pulsars), $\xi$,
$\chi$ and $\sigma_{\rm \xi}$ are variables related to the correlation
function, as defined in \citeasnoun{jhlm05} and $\sigma_{\rm g}$ is
the RMS residuals introduced by the GWB, derived by \citeasnoun{jhlm05}
to be:
\begin{equation}
  \sigma_{\rm g}^2 = \frac{A^2}{12 \pi^2\left(2-2\alpha\right)}
  \left(f_{\rm l}^{2 \alpha -2 }-f_{\rm h}^{2 \alpha - 2}\right)
\end{equation}
with $f_{\rm h} = 2/\Delta t$ and $f_{\rm l}=1/T$ the highest and
lowest detectable GWB frequencies based on the span of the timing
observations $T$, $\Delta t$ the sampling interval, and $A$ and
$\alpha$ the amplitude and spectral index of the GWB in the
characteristic strain spectrum: $h_{\rm c}(f) = A
\left(f/f_0\right)^{\alpha}$ where $h_{\rm c}$ is the characteristic
strain and $f_0 = 1\,$yr$^{-1}$. While GW frequencies below $T^{-1}$
are originally present in the timing data, they are effectively
removed by the fitting process that estimates the timing model
parameters (most importantly the pulse period and spindown) from the
timing residuals.

In the weak-field limit where the GWB power at the lowest frequency,
$f_{\rm l}$, is comparable to the noise power in the pulsar timing
residuals, Equation \ref{eq:S} can be rewritten into a scaling law for
the GWB amplitude to which a PTA is sensitive:
\begin{equation}\label{eq:Scaling}
  A_{\rm GWB} \propto \frac{\sigma_{\rm n}}
  {T^{5/3} \sqrt{N M \left(M-1\right)}}.
\end{equation}
Note that this relation only holds for the GWB originating from
supermassive black hole binaries, which has a predicted spectral index
of $\alpha = -2/3$ \citeaffixed{svc08}{see e.g.}. This negative
spectral index implies that PTAs are most sensitive to GWB frequencies
of the order of the inverse of the data length: $T \approx 10\,$years
or $f_{\rm l} \approx 3\,$nHz.

Equation \ref{eq:Scaling} clearly delineates the four parameters that
are fundamental to the success of a PTA: the number of pulsars $M$,
the number of observing epochs $N$, the RMS of the timing residuals
$\sigma_{\rm n}$ and the length of the data set $T$. Since any one
telescope is limited in the number of available pulsars and the
regularity with which it can observe them, $M$ and $N$ are practically
fixed at around 20\,MSPs and weekly to biweekly
observations. International collaboration, however, has the potential
to significantly improve on these numbers, as described elsewhere in
these proceedings by \citeasnoun{haa+09}.

\section{Recent and ongoing progress}\label{sec:progress}

The previous section showed how PTA sensitivity scales with RMS
residuals and with the length of the timing data sets. In this section
we will investigate how these parameters can be used to improve our
sensitivity to GWs. Specifically, in \S\ref{ssec:technology} we
comment on continuing improvements to observing hardware and data
analysis methods that will decrease receiver noise levels and reduce
some systematic effects that currently contribute to the RMS
residuals, $\sigma_{\rm n}$. Since PTA sensitivity does not simply
depend on $\sigma_{\rm n}$, but rather on $\sigma_{\rm n}/T^{5/3}$, it
is important also to assess how the RMS residuals evolve over
increasing time-spans. For many young pulsars it is known that timing
deteriorates over long time-spans because of irregular spindown or
``timing noise''\cite{antt94}. In \S\ref{ssec:intrinsic} we will
highlight some recent investigations into the presence of such
low-frequency noise in our MSP timing data.

\subsection{Technical improvements}\label{ssec:technology}
One of the cornerstones of the PPTA efforts to improve RMS residuals
values, is improving the observational hardware. Development has been
undertaken along several lines:
\begin{description}
\item[Observing systems:] Two new observing systems have been
  commissioned within the last year. The Parkes Digital Filterbank
  system (PDFB3 and PDFB 4) are a digital polyphase filterbank that
  are capable of 8-bit sampling at a rate of 2\,GHz for an
  observational bandwidth of up to 1\,GHz. They support a range of
  configurations including an online-folding mode with up to 2048
  pulse phase and radio frequency bins, a search mode with up to 8192
  frequency channels and a baseband sampling mode. The second
  newly-commissioned observing system is the ATNF-Parkes-Swinburne
  Recorder (APSR). It receives up to 1\,GHz of baseband data from the
  PDFB3 and performs real-time coherent dedispersion on a dedicated
  16-node, dual quad-core computing cluster, providing real-time system
  diagnostic and observational information through a web-based user
  interface.
\item[Real-time interference mitigation:] A system for real-time
  mitigation of radio-frequency interference (RFI) as described by
  \citeasnoun{khc+05} has been implemented at the Parkes radio
  observatory to work with the PDFB3 during folded and baseband data
  taking.
% \item[Square kilometre array (SKA) pathfinders:] Members of the PPTA
%   are strongly involved in the development of the Australian SKA
%   pathfinder (ASKAP), which is being built and is expected to be
%   operational from 2012 onwards. Whilst the combined collecting area
%   of the ASKAP elements is not too different from that of the Parkes
%   radio telescope, the increased number of observations and improved
%   RFI environment will improve the PPTA's sensitivity. Furthermore,
%   investigations into optimising timing efforts on a large
%   interferometer will be crucial for future GW investigations with the
%   SKA.
\end{description}

Besides development of observational hardware, the PPTA team is
investigating various ways of improving the timing analysis itself:
\begin{description}
\item[Improved timing software:] As part of the PPTA efforts, the
  \textsc{Tempo2} pulsar timing software package was created
  \cite{hem06}. Based on the original \textsc{Tempo} software, it
  implements all known relevant effects down to a precision of 1\,ns,
  which is several orders of magnitude more precise than most previous
  timing software packages. Recently, \textsc{Tempo2} has been
  expanded with GW simulation capabilities \cite{hjl+09}, allowing
  realistic evaluation of timing model fitting effects on GW
  sensitivity.
\item[ISM and solar wind:] The effect of the changing interstellar
  medium density on pulsar timing residuals and an optimal way to
  correct for these variations was published by
  \citeasnoun{yhc+07}. The varying electron density caused
  specifically by the solar wind was investigated in more detail by
  \citeasnoun{yhc+07b}.
\item[Full polarimetric calibration:] Imperfect calibration of the
  instrumental polarisation causes systematic errors in arrival time
  estimation. Calibration errors are mitigated through measurement
  equation modelling \citeaffixed{van04a}{MEM;}, a technique that uses
  observations made over a wide range of parallactic angles to better
  determine the instrumental response. When used in combination with
  matrix template matching \citeaffixed{van06}{MTM;}, which exploits
  the additional timing information in the polarisation of the pulsar
  signal, the RMS residuals of recent data from PSR J0437$-$4715
  is reduced by up to 50\% \cite{vbv+08}.
\item[Improved solar-system parameters:] Pulsar timing has reached a
  precision where accurate solar-system ephemerides crucially affect
  both timing precision and the resulting timing model parameters
  \cite{vbv+08}. This implies that, rather than correcting for the
  position and mass of solar-system bodies, PTA data sets are now able
  to add information to solar-system models. An initial investigation
  of this capability, including a precise measurement of the Jovian
  system mass, is about to be published \cite{cha+09}.
\item[Frequency-dependent template profiles:] The 1\,GHz bandwidth
  that the newly-commissioned observing systems can record, is
  increasing the amount of possible pulse profile evolution across a
  single observing band. Ignoring this issue would artificially
  broaden the pulse profile when integrated over the observed
  frequency range - and therefore worsen TOAs and timing
  residuals. Research on profile evolution and
  observing-frequency--dependent template profiles to time against is
  ongoing.
\item[Pulse stabilisation time-scales:] The shape of the pulses
  received from a radio pulsar, changes from one pulse to the
  next. Luckily, integration of many subsequent pulses results in a
  stable, reproducible profile. With increased bandwidths and future
  timing programs at highly sensitive telescopes such as the SKA, it
  is important to know the precise stabilisation time and therefore
  the shortest pulse-integration time suitable for high-precision
  timing. Using the new APSR backend (see \S\ref{ssec:technology}), we
  have recently commenced such a study of single pulses and
  pulse-stabilisation time-scales.
% \item[Pulse shape stability:] It is conceivable that the shape of
%   pulse profiles may not be constant in time, for a variety of
%   reasons. Geodetic precession in relativistic binary systems may be
%   one cause \citeaffixed{wrt89,atw99,hbo05}{see, e.g. }, free
%   precession may be a cause in isolated pulsars \cite{sls00} and for
%   some of the PPTA pulsars instabilities originating in the pulsar
%   magnetosphere were reported by \citeasnoun{kxc+99}. For PSR
%   J1022+1001, \citeasnoun{hbo04} rejected this claim. Using the new
%   generation of observing systems, we are now undertaking a close
%   inspection of pulse profile stability on all our 20\,MSPs.
\end{description}

\subsection{Intrinsic pulsar properties}\label{ssec:intrinsic}
The improvements outlined in \S\ref{ssec:technology} are guaranteed to
reduce the radiometer noise and systematic contributions to the RMS
residuals. The ultimate potential of PTAs does, however, not depend
exclusively on hardware and processing techniques: eventually, timing
will be limited by the intrinsic characteristics of the pulsars
themselves. There are two specific properties of pulsars that are of
relevance to the future potential of PTAs: the lowest achievable
RMS residuals ($\sigma_{\rm n, min}$) and the `timing stability' of the
pulsars, defined by \citeasnoun{vbc+09} as ``the potential of an MSP
timing data set to maintain a constant, preferably low RMS residuals at
all time-scales up to the time-span of a PTA project, which is
typically envisaged to be 5 years or longer.''

Referring back to Equation \ref{eq:Scaling}, the importance of these
two properties is evident: $\sigma_{\rm n, min}$ uniquely defines the
minimal GWB amplitude to which a PTA with given number of pulsars and
observing cadence can be sensitive to within a given length of
time. The timing stability, on the other hand, specifies if
$\sigma_{\rm n} T^{-5/3}$ decreases as a function of time or not - and
hence if the PTA gains sensitivity because of longer time-spans, or
loses sensitivity because of intrinsic low-frequency instabilities in
the timing data.

\begin{figure}
  \psfig{angle=0.0,width=9cm,figure=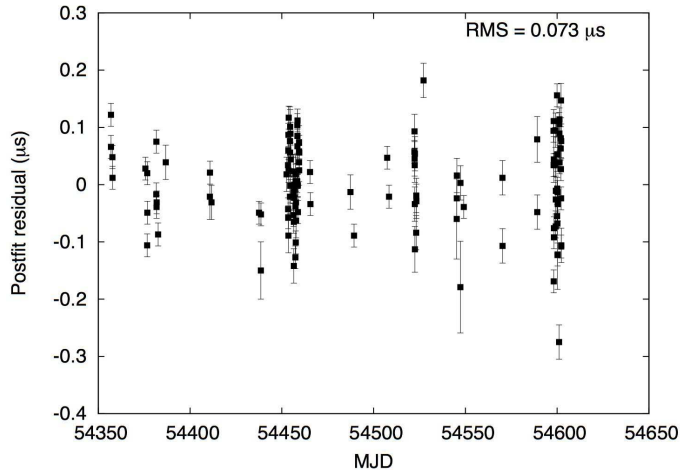}
  \caption{ Timing residuals for PSR J0437$-$4715, obtained with the
    PDFB2 observing system at the Parkes Radio Observatory. The timing
    model of \citeasnoun{vbv+08} was used and only the pulse period
    and spindown rate were refitted.}
  \label{fig:0437res}
\end{figure}

\begin{figure}
  \psfig{angle=0.0,width=9cm,figure=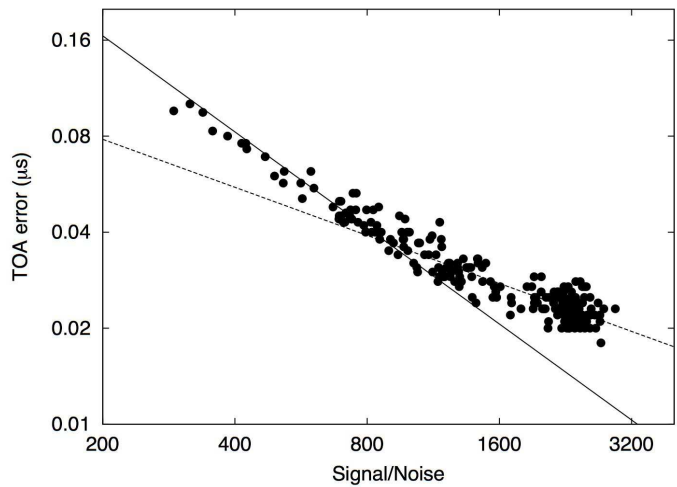}
  \caption{TOA uncertainty versus signal-to-noise ratio (S/N) for
    PDFB2 data on PSR J0437$-$4715 (dots), timed against a
    fully-calibrated, high-S/N CPSR2 (Caltech-Parkes-Swinburne
    Recorder 2) template profile. The full line shows the
    theoretically expected inverse relationship, the dashed line
    scales $\sigma_{\rm TOA}$ as $1/\sqrt{\rm S/N}$. In contrast to
    the similar plot shown by \citeasnoun{hbb+09}, these data
    demonstrate the potential to achieve TOA precisions down to
    20\,ns. The departure from theoretical scaling in the high S/N
    regime suggests the presence of data analysis imperfections such
    as calibration errors, or limits in precision caused by variations
    in observing frequency, for example.}
    %The TOA uncertainty
    %mostly follows the theoretical scaling with S/N, though
    %corruptions flatten this relation in the very high-S/N regime,
    %possibly due to calibration errors or data analysis
    %imperfections.}
  \label{fig:hbb}
\end{figure}

An upper bound on $\sigma_{\rm n,min}$ can most readily be determined
as the lowest RMS residuals obtained so far. Initial data from the
PDFB2 pulsar backend (the immediate precursor to the PDFB3 mentioned
in \S\ref{ssec:technology}) on PSR J0437$-$4715 in the $\nu\sim3,$GHz
observing band, have already achieved weighted RMS residuals of 73\,ns
over 0.7\,years, with 126\,TOAs (see Figure \ref{fig:0437res}). A
somewhat more involved analysis \cite{vbc+09} that used
multi-frequency data to separate out several contributions to the RMS
residuals, established a bound of 80\,ns on a time-scale of 5\,years
for PSRs J1909$-$3744 and J1713+0747. A third type of analysis was
presented by \citeasnoun{hbb+09} who investigated how the TOA
uncertainty varies with the signal-to-noise ratio (S/N) of the
observation. While this approach is only sensitive to a subset of
contributions to the RMS residuals, it does have the ability to
highlight systematic problems in the data analysis. In their Figure 2,
\citeasnoun{hbb+09} demonstrated how TOA uncertainties grew
dramatically worse for S/Ns above $\sim$1000, suggesting systematic
effect at levels up to 100\,ns. We present a re-analysis of their data
in Figure \ref{fig:hbb}, improving the pulse template used in the
cross-correlation from which the TOA uncertainty is derived. Our
results still contain systematic worsening of the TOA uncertainties at
high S/N, but these effects now only occur at the $20-30$\,ns level.

\begin{figure*}
  \psfig{angle=0.0,width=11cm,figure=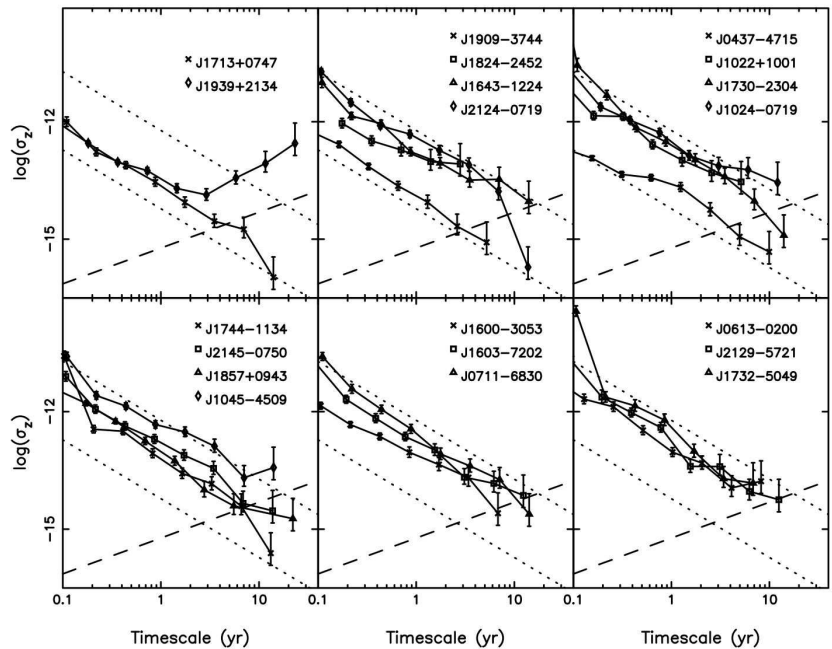}
  \caption{$\sigma_{\rm z}$ stability curves for the 20\,MSPs that
    constitute the PPTA, as in \citeasnoun{vbc+09}. The downward
    sloping, dotted lines are theoretical white noise levels at
    100\,ns (lower line) and 1\,$\mu$s (upper line). The upward
    sloping, dashed line is a theoretical curve for a hypothetical
    GWB. These graphs demonstrate the large excess of low-frequency
    noise in the PSR J1939+2134 data set, which could easily obscure a
    GWB. For some other pulsars (PSR J1045$-$4509, for example) there
    is some indication of low-frequency noise, though more data would
    be required to verify this. Overall, the lack of clear
    low-frequency noise bodes well for long-term GW detection
    efforts.}
  % For all other pulsars, the graph shows no significant low-frequency
  % noise, though some pulsars (such as PSR J1045-4509) which bodes well
  % for long-term GW detection efforts.}
  \label{fig:Stab}
\end{figure*}

\citeasnoun{vbc+09} also investigated the long-term timing behaviour
of the PPTA pulsars based on the $\sigma_{\rm z}$ stability parameter
defined by \citeasnoun{mte97}: $\sigma_{\rm z}(\tau) = \frac{\tau^2}{2
  \sqrt{5}}\sqrt{\langle c^2_3\rangle}$, with $\tau$ the timescale
considered, $c_3$ the amplitude of a third-order polynomial fitted to
a subset of the timing residuals of length $\tau$, and $\langle\
\rangle$ denotes averaging over different subsets if $\tau$ is less
than the time-span of the observations. Figure \ref{fig:Stab}
summarises their results: on a 5\,yr time-scale only PSR J1939+2134
shows non-white noise at high enough levels to obscure a GWB. On
longer time-scales ($\sim 10$\,yr) the data also show potential
low-frequency excess power shows in some other pulsars. For most
pulsars in our sample it therefore appears likely that the GWB signal
will dominate our timing on time-scales between 5 and 10\,years,
provided timing residuals are decreased further. While detection would
be based on cross-correlations of timing residuals instead of the
timing residuals of a single pulsar, unmodelled low-frequency noise as
present in the PSR J1939+2134 data would render any GW-induced
correlation less significant.
%of all 20\,MSPs, only PSR
%J1939+2134 shows 
%a clear deviation from white noise on a time-scale of
%$\sim5$\,yr. %has sufficiently steep intrinsic noise at a level high
%enough to obscure the influence of a GWB that might be present in the
%data on time-scales of years to decades. It must be noted that PSR
%J1824$-$2452 can also be expected to have high levels of low-frequency
%noise due to the accelerations it experiences in the gravitational
%potential of its globular cluster. However, the data set on PSR
%J1824$-$2452 analysed by \citeasnoun{vbc+09} is only 2.8\,years long,
%which implies the low-frequency noise is not yet sufficiently sampled.

\section{Sensitivity to GWs}\label{sec:GWBSens}
There are two main classes of GW sources that may be detectable by
PTAs: GWBs and single sources of GWs. In both cases the best studied
sources are super-massive black-holes binary systems (SMBHBs): with low
enough RMS timing residuals, nearby SMBHBs may be strong enough to be
detected individually and the combined effect of a large number of
weaker, more distant ($z \approx 1 - 2$) SMBHBs is expected to be
detectable as a GWB. In the following section, the PPTA's sensitivity
to single sources of SMBHBs will briefly be discussed, while
projections of sensitivity to GWBs will be described in
\S\ref{ssec:vbc+09}, assessing realistic time-scales for a positive
detection of the GWB.

\subsection{Sensitivity to Single Sources of GWs}\label{ssec:Yardley}
The long-term timing of the PPTA pulsars \cite{vbc+09} was recently
used by \citeasnoun{yhj+09} to determine the sensitivity of the PPTA
to GWs emitted by SMBHBs. The conclusion of this research is that
most\footnote{Because of the need to fit the parameters of the pulsar
  timing model, the sensitivity at yearly and half-yearly periods is
  much reduced.}  $10^{10}\,M_{\odot}$ SMBHBs with orbital frequencies
of order $10^{-9}$ to $10^{-6}$\,Hz at distances out to the Virgo
cluster can already be excluded, with sensitivity falling sharply with
distance. They further predict that the high-mass ($\sim 5\times
10^9\,M_{\odot}$), high-redshift ($z \geq 2$) end of SMBHB merger
rates will be constrained by pulsar timing within a decade.

\subsection{Predicted GWB Detection Time-scale}\label{ssec:vbc+09}
Using the same long-term data sets and building on the analysis of
\citeasnoun{jhlm05}, \citeasnoun{vbc+09} constructed the first GWB
sensitivity curve for a PTA with a realistic spread in RMS residuals for
the different pulsars. That sensitivity curve showed a $3\,\sigma$
detection sensitivity for GWB amplitudes $A \geq 2\times 10^{-14}$,
nearly equal to the best limit on the GWB amplitude: $A < 1.1\times
10^{-14}$ \cite{jhv+06}. Assuming a conservative noise floor of
$\sigma_{\rm n,min}^{\rm est} = 80\,$ns, \citeasnoun{vbc+09}
subsequently scaled their RMS residuals values with telescope
sensitivity to make straightforward predictions for the sensitivity of
other PTA efforts and projected the PPTA's sensitivity on a decadal
time-scale. They concluded that within another five to ten years, the
PPTA is likely to obtain a $3\,\sigma$ sensitivity level throughout
most of the predicted amplitude range of GWBs composed of
SMBHBs. While this analysis does not predict detection significances
beyond $3\,\sigma$, this is mainly due to the bound of $\sigma_{\rm
  n,min}^{\rm est}$, which may be too restrictive, especially in the
light of the material presented in \S\ref{ssec:intrinsic}.

\section{Conclusions}\label{sec:Conc}
We have provided an overview of recent and ongoing work involved with
the Parkes Pulsar Timing Array, focussing on hardware and software
development and on investigations of fundamental pulsar properties
that may be crucial to an eventual PTA-based detection of GWs. We have
presented results of sensitivity calculations for the current PPTA
data set to both single sources and a background of GWs. Making
reasonable assumptions about the continued improvement of our data
quality, we predict that the PPTA has a reasonable chance of detecting
either the GWB or the GWs from a single, nearby SMBHB system within
the next decade. This sensitivity will only be helped by various other
projects, such as ongoing and planned pulsar surveys, the
international PTA collaboration described elsewhere in these
proceedings \cite{haa+09} and new, highly sensitive telescopes such as
the SKA and its pathfinder telescopes.

\ack The Parkes Observatory is part of the Australia Telescope which
is funded by the Commonwealth of Australia for operation as a National
Facility managed by CSIRO. We thank the staff at Parkes Observatory
for technical assistance and dedicated help and we acknowledge the
dedication and skills of the engineers involved in this project. JPWV
acknowledges support from a WVEPSCoR research challenge grant held by
the WVU Center for Astrophysics.

\section*{References}
\bibliographystyle{jphysicsB}
\bibliography{journals,psrrefs,modrefs,crossrefs}

\end{document}